# The Orion constellation as an installation

An innovative three dimensional teaching and learning environment

*Dr. Daniel Brown* School of Science and Technology, Nottingham Trent University

Visualising the three dimensional distribution of stars within a constellation is highly challenging for both students and educators, but when carried out in an interactive collaborative way it can create an ideal environment to explore common misconceptions about size and scale within astronomy. We present how the common table top activities based upon the Orion constellation miss out on this opportunity. Transformed into a walk-through Orion installation that includes the position of our Solar system, it allows the students to fully immerse themselves within the model and experience parallax. It enables participants to explore within the installation many other aspects of astronomy relating to sky culture, stellar evolution, and stellar sizes establishing an innovative learning and teaching environment.

## I. INTRODUCTION

Stars and other astronomical objects are typically mapped on the surface of an imaginary celestial sphere with only the direction towards the objects recorded. However, this has led to a common misconception that stars within a constellation are at the same distance, thereby ignoring that constellations are mere imaginary figures or defining boundaries structuring the sky. In the 20$^{th}$ century the International Astronomical Union (IAU) decided to define 88 constellations as regions each containing the traditional constellation and all the objects in the area. [1]

Educational research into how the structure of the solar system is perceived by students of different age has concluded that stars are quite commonly placed within our solar system.[2] These problems originate from the challenge of imagining the great distances towards some stars. Even though these misconceptions are slowly corrected, they are still present in trainee science teachers. Brown & Neale have found evidence that there is a problem in fully understanding the sheer distance of the stars from us.[3] Their findings highlight the misinterpretation of parallax and essentially of the scale of the distance travelled on Earth with respect to the distances towards the stars in constellations; as a consequence there is a need for a more focused activity targeting astronomical size and scale within schools similar to examples of human orrerys [4].

This article will introduce a simple but versatile activity visualising the three dimensional distribution of stars in the constellation of Orion. This constellation is one of the most commonly known and recognised ones. There are several good examples of the distribution of the stars in Orion using virtual models [5], small table tops [6,7], and small classroom activities [8,9]. However, many are small scale models that do not allow for an environment in which collaborative learning can take place. These models are so compact that a critical engagement with scale becomes very abstract reducing the impact of the model.[10] There are some cases in which the constellation is established as a group activity but the outcome can only be explored afterwards limiting the potential of collaborative work. The most important short-coming of all non-virtual models is the neglect of the location of our solar system. Without a reference point such models become abstract and difficult to engage with.

The following activity has a more immersive approach offering a joint exploration of a larger installation that includes the actual location of our solar system. Such an installation enables the students to discover that some of the stars in the Orion constellation exceed the distance of the closest star of Orion to our solar system and offers sufficient space to carry out discussions and activities within the actual model.

Table I lists the coordinates of the objects used in the Orion constellation.

Table I. The right ascension($\alpha$) and declination ($\delta$) and distance ($r$) coordinates towards the main objects in the Orion constellation are listed. The Earth is located at (0,0,1) and 1 cm is equivalent to 1 light year.

| # | Object | $\alpha$ [h:m] | $\delta$ [d:m] | $r$[ly] | $x$[m] | $y$[m] | $z$[m] |
|---|--------|-------|-------|------|-------|-------|-------|
| 1 | alpha | 5:55 | 7:24 | 427 | -0.41 | 4.12 | 2.06 |
| 2 | beta | 5:15 | -8:11 | 773 | 0.57 | 7.71 | 0.84 |
| 3 | gamma | 5:25 | 6:22 | 243 | 0.07 | 2.36 | 1.56 |
| 4 | delta | 5:32 | 0:17 | 916 | 0.00 | 9.10 | 2.07 |
| 5 | epsilon | 5:37 | -1:11 | 1342 | -0.29 | 13.35 | 2.36 |
| 6 | zeta | 5:41 | -1:56 | 817 | -0.32 | 8.13 | 1.72 |
| 7 | kappa | 5:48 | -9:40 | 722 | -0.50 | 7.19 | 0.66 |
| 8 | iota | 5:36 | -5:54 | 1325 | -0.23 | 13.25 | 1.25 |
| 9 | M42 | 5:36 | -5:23 | 1344 | -0.23 | 13.43 | 1.38 |

## II. Orion Installation

It is assumes the teacher has an outdoor space of sufficient size available. Column $z$ in Table I marks the length of the poles above ground. The poles need to be topped by star globes. Make sure to place a pole of height $z_0 = 1$ m at the origin. This pole will mark the location of the solar system and should remain empty to ease observations and invited the students to find something that could represent the size of our solar system. It is highly recommended to involve the students in all steps of the building of this installation.

## III. Exploring Orion Installation

From the solar system pole, the students can see the constellation of Orion, as is familiar from Earth, however it will rapidly be distorted from any other position allowing for a later follow up on changes of constellations in space and time. The scale is chosen so that 1cm represents 1 light year.

To revisit their knowledge of our Sun-Moon-Earth system, the teacher should now ask the students to place themselves at the distance where they think the Moon would be in our model. Experience has shown that they will choose many different distance throughout the installation.

To develop an understanding of the real depth of space represented in the installation, the students should be asked to return and to look for something in the outdoor space that could represent the size of our entire solar system in our installation. The teacher can reinforce the scale by pointing out that the distance of the closest star system in the installation is 5 cm.

Possible follow up:
- Changing constellations in space and time
- Sizes of stars
- Light travel times and light years
- Stellar evolution

## IV. CONCLUSION

Exploring the vast distances and sizes involved in astronomy is challenging and many table top models visualising the distribution of stars in Orion struggle making the distances involved fully comprehendible. We have presented an innovative and creative way of building the Orion constellation that engages students in the challenging topic of distances in astronomy and thereby extending already good practice regarding solar system models. The installation presented here allows for a fully collaborative and interactive learning environment suitable for a range of student abilities. We have described a route of different learning experiences offering possibilities to differentiate and cater towards all the different learning styles within a typical class.

## ACKNOWLDGEMENTS

D.B. acknowledges the financial support of IGNITE! in developing the Orion installation together with F. Johnston at the Beaumont Leys School, Leicester, as well as R. Francis and the Nottingham Contemporary. D. B. also wishes to thank R. Doran for her support in writing this article.

## REFERENCES

[1] IAU, 2011. The Constellations [online]. IAU. Available at: <http://www.iau.org/public/constellations/> [Accessed 15 October 2011].

[2] R., Trumper, "The need for change in elementary school teacher training--a cross-college age study of future teachers' conceptions of basic astronomy concepts", Teaching and Teacher Education 19, 309-323 (2003).

[3] D., Brown, & N., Neale, "A Global Citizen of the Skies", Educational futures (BESA) 2 (2), 41-55 (2010).

[4] P. Newbury, "Exploring the Solar System with a Human Orrery" The Physics Teach. 48, 573 (2010)

[5] V. Kuo and R Beichner, "Stars of the Big Dipper: A 3-D vector activity," The Physics Teacher 44 (4), 168-172 (2006).

[6] A., Breidenbach, 2002, Bastelbogen "Orion" [online]. Sternwarte Recklinghausen. Available at: <http://www.sternwarte-recklinghausen.de/files/orion.pdf> [Accessed 15 October 2011].

[7] D., MacDonald, & A., Manuel, 2007, Teacher Worksheets, Orion in 3D. Allsky.ca. Available at: <http://allsky.ca/worksheets/Orion%20in%203D.pdf> [Accessed 15 October 2011].

[8] J. A., Navarro, & J. M., Rodríguez, 2011, LA CONSTELACIÓN DE ORIÓN [online], Network for Astronomy School Education UNAWE. Available at: <http://sac.csic.es/unawe/Actividades/CONSTELACIONDEORION.pdf> [Accessed 15 October 2011].

[9] R. M., Ros, A., Capell, J., Colom, 2011, CONSTELACIONES EN 3D [online], Network for Astronomy School Education UNAWE. Available at: <http://sac.csic.es/unawe/Actividades/constelaciones%20en%203D.pdf> [Accessed 15 October 2011].

[10] E. Etkina, A Warren, and M. Gentile, "The Role of Models in Physics Instruction", The Physics Teacher 44, 34 (2006)